\documentstyle[sprocl,epsfig]{article}

\def\met{\mbox{$\not\!\!{E}_{T}\,$}}
\def\et{\mbox{$E_T$}}
\def\pt{\mbox{$p_T$}}

\def\et{\mbox{$E_T$}}
\def\pt{\mbox{$p_T$}}

\def\gev{{\rm~GeV}}

\def\met{\mbox{$\not\!\!{E}_{T}$}}

\def\dinkyLQ{{\scriptscriptstyle LQ}}
\def\dinkyTC{{\scriptscriptstyle TC}}
\def\dinkyT{{\scriptscriptstyle T}}

\begin{document}
\begin{flushright}
Fermilab-Conf-99/013-E
\end{flushright}

\title{Tests of the Electroweak Symmetry Breaking Sector
at the Tevatron
\footnote{Lecture given at the XXVIth International Meeting on 
Fundamental Physics, La Toja, Spain, June 1998}}

\author{J.F. DE TROCONIZ
\footnote{Representing the CDF Collaboration}
}

\address{Dpto. de F\'{\i}sica Te\'orica\\
Universidad Aut\'onoma de Madrid,\\
E28049 Cantoblanco, Spain}

\maketitle

\abstracts{
The phenomenology of the electroweak symmetry breaking sector is expected
to involve resonances decaying preferentially into heavy fermions. 
Three recent CDF analyses are reviewed and used to constrain Higgs and 
Technicolor models.}

\section{Introduction}
For many years the standard model has been remarkably successful at explaining
and predicting experimental data. However the details of the electroweak 
symmetry breaking (EWSB) sector of the theory remain largely unknown, and 
one of the primary goals of present and future colliders is to uncover 
the mechanism responsible for the symmetry breaking.

In the standard model and many extensions to it, 
the electroweak symmetry is spontaneously broken 
by introducing fundamental scalar particles into the theory.
These are eventually identified with $W_L$, $Z_L$, and 
one or more physical Higgs bosons~\cite{theo2}.
In fact the standard model incorporates the simplest implementation,
with just one scalar field doublet.
This leaves a single observable scalar particle, the Higgs boson,
with unknown mass but fixed couplings to other particles.  
It should be noted that this simple model explains several striking facts,
and any complication (even if possible) tends to weaken it~\cite{veltman}.

Alternatively, the electroweak symmetry may be broken dynamically.
This is the hallmark of technicolor (TC) theories \cite{tc} in which
a new strong gauge force (technicolor) and new fermions (technifermions)
are introduced.  The technicolor force is modelled after QCD, 
scaled up to the TeV scale, with the technifermions being the analogs of 
ordinary quarks. Technicolor acts between the technifermions to form bound
states (technihadrons). In particular, the technipions include the 
longitudinal weak bosons, $W_L$ and $Z_L$, as well as the pseudo-Goldstone 
bosons of dynamical symmetry breaking. Thus the dynamics of the 
technifermions assume the role of the scalar Higgs fields in theories with 
spontaneous symmetry breaking.

Both models coincide in that they can be extended to generate all masses
in the theory, notably the fermion masses. In the first case, explicit 
Yukawa couplings are used. In the second, the first TC model was augmented 
by Extended Technicolor (ETC)~\cite{etc}. In fact, it is attractive to 
assume that a 
connection between the physics of the EWSB and the physics of flavour 
exists even more generally than the two previous examples.
The important point is that both models now predict that the scalars
(fundamental or composite) will have couplings to ordinary fermions that
are proportional to mass, i.e., these scalars will decay predominantly
in channels involving heavy quarks and heavy leptons. 
The experimental consequence of this is to provide a general frame to look 
for the particles involved in the EWSB, namely, to look for resonances 
decaying into heavy quarks and/or heavy leptons produced in pairs or in 
association with a $W$ or $Z$.

In the following we describe three experimental searches recently published 
by the CDF Collaboration using $\sim 100$ pb$^{-1}$ of $p \bar{p}$ collisions 
at the Tevatron~\cite{higgs1,higgs2,lq3}:
\begin{itemize}
\item[(1)] Search for $X \to b\bar b$ in association with $W \to \ell \nu$.
\item[(2)] Search for $X \to b\bar b$ in association with $V \to q\bar q$,
where $V = W, Z$. 
\item[(3)] Search for events including a $\tau^+ \tau^-$ pair and two extra 
jets.
\end{itemize}

No signal beyond standard model expectations has been observed in the
current data sample. So in the rest of the article we will investigate
the implications of this null result on the models cited at the beginning of
the section, emphasizing the first-hand information on the performance of the 
different experimental techniques and direct evaluation of involved 
backgrounds. In particular we use searches (1) and (2) to constrain the
production of a light standard model Higgs, and search (3) to investigate
technicolor models containing a technifamily.

There are other analyses not discussed here for lack of space that can be used
to find information about the EWSB sector: top physics, charged Higgs searches,
etc. The reader will find information in references~\cite{d0}.

\section{Light Standard Model Higgs}

The possible range for the mass of the standard model Higgs extends from a 
lower bound of about 88 GeV$/c^2$ from the LEP experiments~\cite{lep} to 
${\cal O}$(1) TeV. The present analysis has to be restricted to light masses 
for two reasons:
\begin{itemize}
\item The production cross sections at the Tevatron are small and decrease 
rapidly as a function of $M_{H^0}$. In $p\bar{p}$ collisions at
$\sqrt{s}=1.8$ TeV, the Higgs production mechanism with the most 
promising detection possibilities
is $p\bar{p}\rightarrow V + H^0$, with $V = W, Z$. In the framework of 
the standard model, the production cross section in this channel is 
1.3 to 0.11 pb for Higgs masses between 70 to 140 GeV/$c^2$~\cite{theo1}.  

\item The technique used is to look for resonances in the $b\bar b$ channel.
$H^0\rightarrow b\bar{b}$ is the dominant decay mode of the Higgs boson
only up to  $M_{H^0} \sim$ 130 GeV/$c^2$. 
\end{itemize}

However there are two important motivations to emphasize the region of light
Higgs masses:
\begin{itemize}
\item Precision electroweak experiments suggest that the Higgs 
boson mass may lie at the lower end of the open range~\cite{wards}. 

\item A similar symmetry breaking mechanism occurs in the minimal
supersymmetric extension of the standard model, where several observable
scalar states are predicted, the lightest of which is expected to 
decay predominantly into $b\bar b$ states and to have a mass below 
135 GeV$/c^2$~\cite{acco}. 
\end{itemize}

The cross sections expected in the standard model are out of the scope of the 
present analysis, using $\sim 100$ pb$^{-1}$ of $p\bar{p}$ collisions.
We therefore report on a general search for a Higgs scalar 
produced in association with a vector boson 
with unknown cross section $\sigma_{V H^{0}}$. 
We look for $H^0$ decays to a $b\bar{b}$ pair with unknown branching ratio 
$\beta$, and for two possible decays of the vector boson: ($i$)
$W \rightarrow \ell \nu$ with $\ell = e, \mu$ and ($ii$) $V
\rightarrow q\bar{q}$. Finally, the limits obtained in both channels are 
combined.

\subsection{Leptonic Analysis}

The experimental signature considered is $WH^0$ with $W \to e \nu$ or 
$\mu \nu$,
and $H^0 \to b\bar b$, giving final states with one high-$p_T$ lepton, large
missing transverse energy ($\met$) due to the undetected neutrino, and two $b$ 
jets~\cite{higgs1}. The ability to tag $b$ jets with high efficiency and a low 
mistag rate is vital for searching for the decay of $H^0 \to b\bar b$. We use
the secondary vertex (SECVTX) and soft-lepton (SLT) $b$-tagging algorithms
developed for the top quark discovery~\cite{secvtx}.

\begin{small}
\begin{table}[t]
\begin{center}
\caption{The expected number of tagged $W + n$ jet events from 
QCD backgrounds and top production compared the observed number of events.}
\label{lt1}
\begin{tabular}{ccccc}
\hline
\hline
 & $W + 1$ jet & $W + 2$ jet & $W + 3$ jet & $W + \ge 4$ jet \\
\hline
\multicolumn{5}{c}{Events with single tag}\\
Expected & $67 \pm 13$ & $ 30 \pm 5$ & $16 \pm 2$ & $13 \pm 3$ \\
Observed & 66 & 36 & 11 & 12 \\
\hline
\multicolumn{5}{c}{Events with double tag}\\
Expected & 0 & $ 3.0 \pm 0.6$ & $3.6 \pm 0.6$ & $5.2 \pm 1.1$ \\
Observed & 0 & 6 & 6 & 2 \\
\hline
\hline
\end{tabular}
\end{center}
\end{table} 
\end{small}

  \begin{figure}[t]
    \centerline{
    \hskip -0.7cm
    \epsfig{file=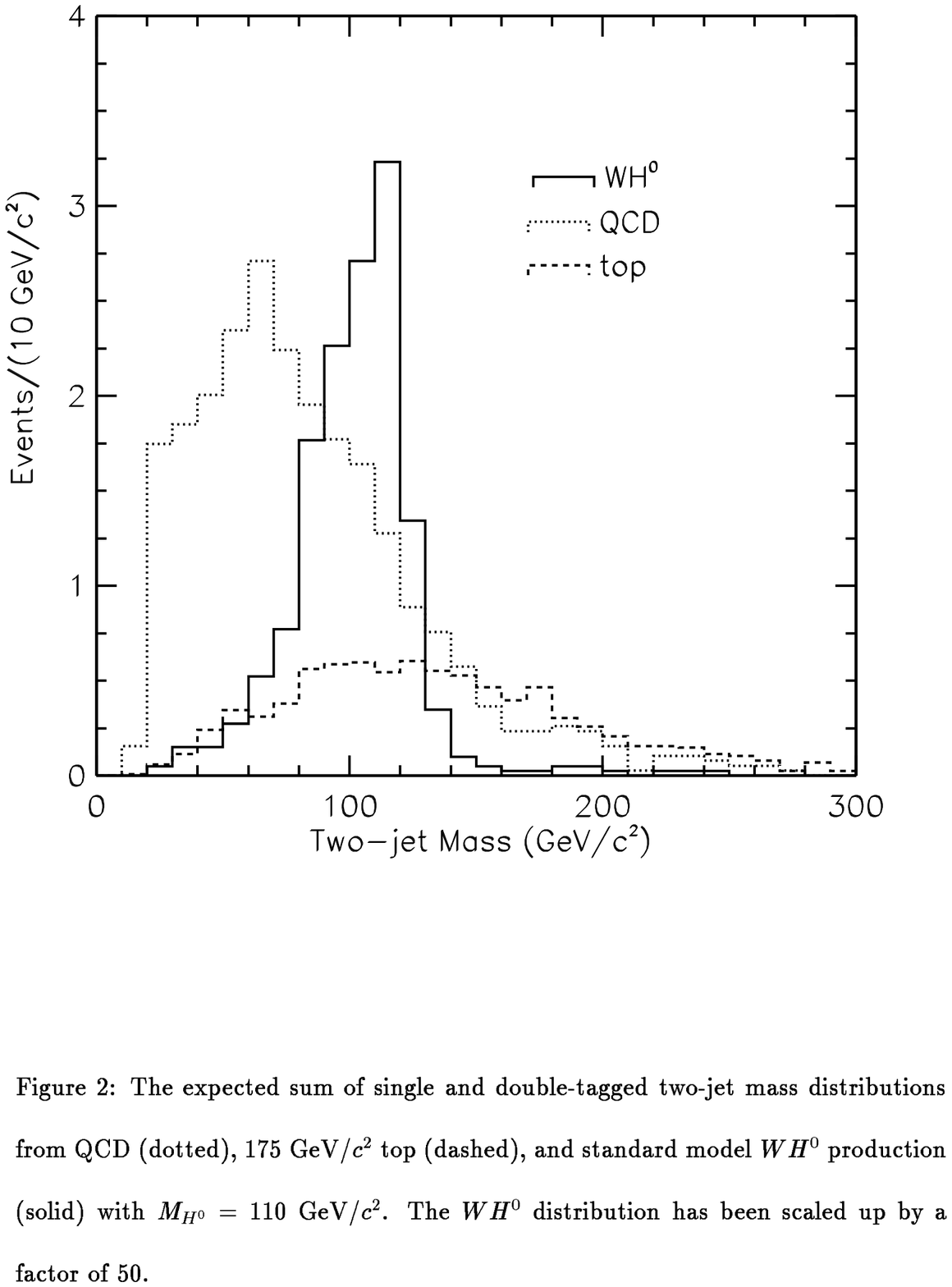,bbllx=60pt,bblly=206pt,bburx=515,bbury=635pt,
                        width=7.8cm,clip=}
    \hskip -0.5cm
    \epsfig{file=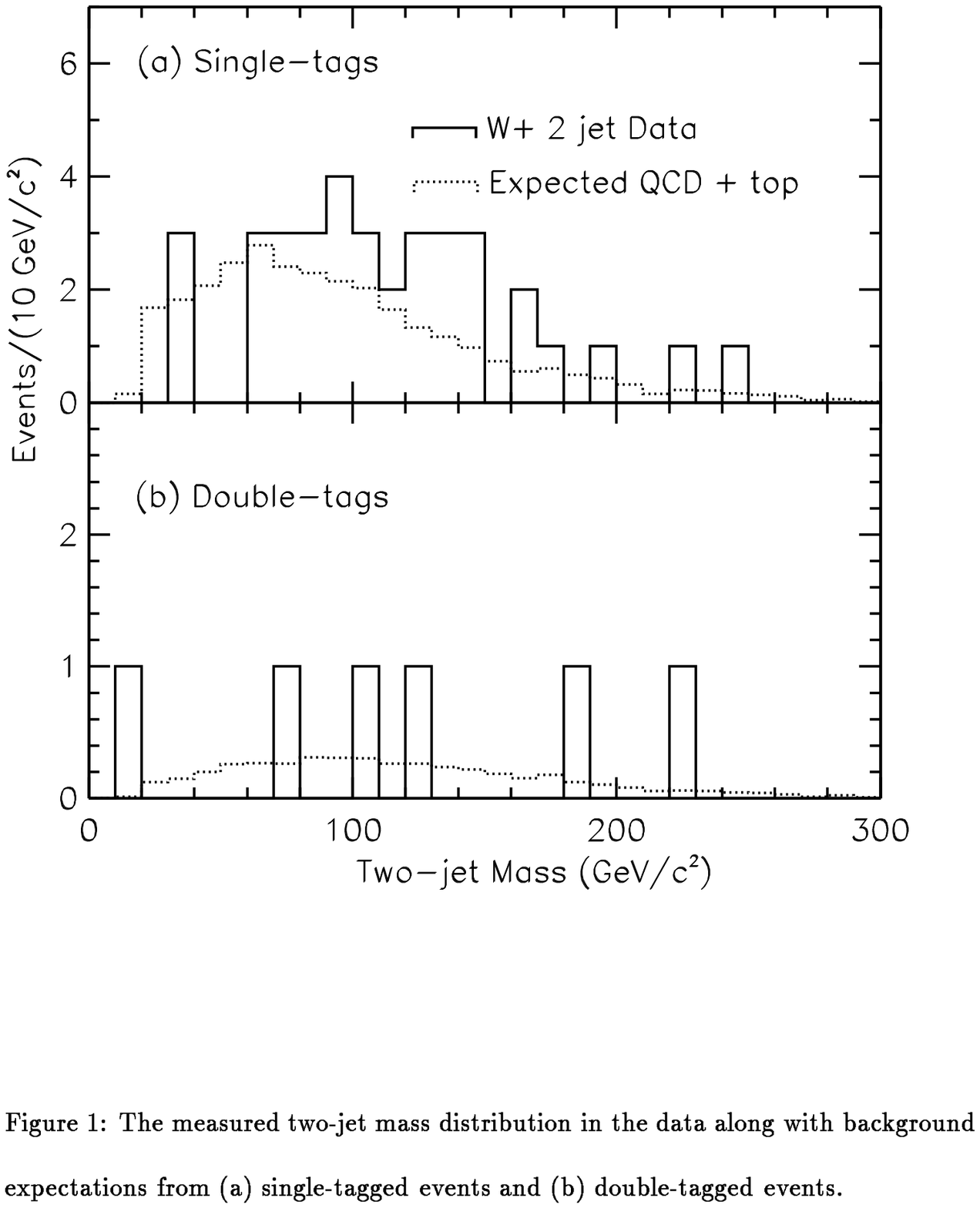,bbllx=60pt,bblly=206pt,bburx=515,bbury=635pt,
                        width=7.8cm,clip=}}
    \caption{(Left) 
The expected tagged two-jet mass distribution from QCD (dotted),
top (dashed), and $WH^0$ production (solid) with $M_{H^0} = 110$ GeV$/c^2$.
The $WH^0$ distribution has been scaled up by a factor of 50.
             (Right) 
The measured two-jet mass distribution in the data along with
background expectations from (a) single-tagged events and (b) double-tagged
events.}
    \label{lf1}
  \end{figure}

A three-level trigger selects events that contain an electron or muon for
this analysis. The event selection starts with the requirement of a primary 
lepton, either an isolated electron with $E_T > 20$ GeV~\cite{CDFcoo} or an 
isolated muon with $p_T > 20$ GeV$/c$, in the central region $|\eta| < 1.0$. A 
$W$ boson sample is selected by requiring $\met > 20$ GeV. Events which 
contain a second, same flavour lepton with $p_T > 10$ GeV$/c$ are removed as 
possible $Z$ boson candidates if the reconstructed $ee$ or $\mu\mu$ invariant 
mass is between 75 and 105 GeV$/c^2$. The events must also not be accepted by
the CDF top dilepton analysis~\cite{secvtx}. To further reduce the dilepton 
backgrounds, we reject events with an additional high-$p_T$ isolated track
with opposite charge to that of the primary lepton. The remaining
events are classified according to jet multiplicity. 
Jets are defined as localized 
energy depositions in the calorimeters and are reconstructed 
using an iterative clustering algorithm with a fixed cone of radius
$\Delta R = \sqrt{\Delta\eta^2 + \Delta\phi^2} = 0.4$ in 
$\eta - \phi$ space~\cite{jets}. Jet e\-ner\-gies 
are then corrected for energy losses in uninstrumented detector regions, 
energy falling out\-side the clustering cone, contributions from underlying 
event and multiple interactions, and calorimeter nonlinearities.

The $W + 2$ jet bin
is expected to contain most of the signal, while the other bins are used to
check the background calculation.
In order to enhance the signal in the $W + 2$ jet bin, we require that one or 
both of the jets be identified ('tagged') as coming from a $b$ hadron.
We require at least one jet to be tagged by the SECVTX algorithm, which has
a higher signal-to-noise ratio than the SLT algorithm. For the single-tag
analysis, the other jet must not be tagged, while for the double-tag
analysis the second jet must be tagged by either SECVTX or SLT.

The SECVTX tagging algorithm begins by searching for secondary vertices that
contain three or more displaced tracks. If none are found, the algorithm
searches for two-track vertices using more stringent track criteria. A jet
is tagged if the secondary vertex transverse displacement from the primary
one exceeds three times its uncertainty.

The SLT tagging algorithm identifies electrons and muon from semileptonic 
$b$ decays by matching tracks with $p_T > 2$ GeV$/c$ with clusters of 
electromagnetic energy in the calorimeters or tracks in the muon chambers.
To gain additional background rejection, we require the track to lie 
within a cone $\Delta R < 0.4$ of the axis of a jet and to be displaced in the
transverse plane from the primary vertex by at least two standard deviations
in the jet direction. This latter requirement reduces lepton 
misidentifications by a factor of five while retaining 65\% of the efficiency.

The total acceptance is calculated as a product of the kinematic and geometric
acceptance, trigger, lepton identification, and $b$-tagging efficiencies,
and the $W$ leptonic branching ratios.
The acceptance for identifying $WH^0$ is calculated from data and a standard
model simulation of Higgs production, where the Higgs is forced to decay into
$b\bar b$ with 100\% branching ratio. The acceptance increases monotonically
from $0.53 \pm 0.13\%$ ($0.17\pm 0.04\%$) to $1.1 \pm 0.3\%$ ($0.42\pm 0.11\%$)
for single (double) tagging as $M_{H^0}$ increases from 70 to 120 GeV$/c^2$.
A 25\% systematic in the acceptance comes from uncertainties in the
modeling of initial and final state radiation, jet energy, and $b$-tag,
trigger, and lepton identification efficiencies.

Background events come predominantly from the direct production of $W$ bosons
in association with heavy quarks (estimated using the HERWIG Monte Carlo
 program), mistags (from generic jet data), and $t\bar t$ production 
(normalized to the CDF measured cross section
$\sigma_{t\bar{t}} = 7.6^{+1.8}_{-1.5}$ pb~\cite{pisa}). 
Other small backgrounds are estimated from a combination of Monte Carlo 
simulations and data.

The numbers of observed single-tagged and double-tagged events and the
corresponding background estimates are shown in Table~\ref{lt1}. By 
construction, data and expectations are in reasonably good agreement
in the $W + \ge 3$ jet bins, which, along with other $t\bar t$ decay
channels, were used to measure the $t\bar t$ production cross section.
The number of $b$-tags in the $W + 2$ jet bin can be compared to the
background calculation. This bin shows a small excess of events 
corresponding to one standard deviation.

To increase the sensitivity of the search we look for a resonant mass peak
in the reconstructed two-jet invariant mass distribution using the 4-momenta
of the jets as measured by the calorimeter.
The expected two-jet invariant mass shape for $WH^0$ production is
shown in Figure~\ref{lf1} (left) for $M_{H^0} = 110$ GeV$/c^2$.
The distributions for the data are shown in Figure~\ref{lf1} (right), 
along with the background expectation.

We set an upper limit on the production cross section times branching ratio
of $p\bar p \to W H^0$ as a function of $M_{H^0}$, by fitting the number of
events in the $W + 2$ jet samples and the shape of the two-jet mass
distributions. The fit yields $\sigma_{W H^0} \cdot {\cal B}(H^0 \to b\bar b)$
in the range from $0.2^{+4.7}_{-0.0}$ to $5.7^{+4.2}_{-3.0}$ pb for a new
particle of mass between 70 and 120 GeV$/c^2$, statistically compatible
with no signal. 
From the 95\% C.L. limits on $\sigma(p\bar{p}\rightarrow WH^0)$,
the corresponding limits on $VH^0$ production were calculated.
We used the program PYTHIA to compute the standard model prediction
for the ratio $\sigma(ZH^0)/\sigma(WH^0)$.
The leptonic analysis efficiency for $ZH^0$ events relative to that for $WH^0$
events was estimated to be $(10\pm 2)$\%.
The limits are summarized in Table~\ref{tab1}.

\subsection{Hadronic Analysis}
The experimental signature considered is 
four jets in the final state, with two of them identified as $b$ 
jets~\cite{higgs2}. The hadronic
channel described here has the advantage of a larger branching ratio,
and is sensitive to both $WH^0$ and $ZH^0$ production
($\sigma(ZH^0)/\sigma(WH^0) \sim 0.6$),
but suffers from a larger QCD background.

\begin{small}
\begin{table}[t]
\begin{center}
\caption{Summary of the hadronic analysis fit results,
standard model predictions for $\beta\sigma$, and 95\% C.L. limits 
from the hadronic, leptonic, and combined analyses.}
\label{tab1}
\begin{tabular}{cccccc}
\hline
\hline
$M_{H^0}$ & $\beta\sigma$ (pb) & $\beta \sigma$ (pb) &
$\beta \sigma$ (pb) & 
$\beta\sigma$ (pb) & $\beta\sigma$ (pb) \\ 
(GeV/$c^2$) & fit & SM & had. limit  & lep. limit & 
comb. limit \\ [0.1cm] \hline
 & & & & & \\ [-0.28cm]
$ 70$  & $44 \pm 42$ & $1.13$ &
   $117.3$ & $21.9$ & $23.1$ \\[0.1 cm] 
$ 80$  & $0^{+19}_{-0}$ & $0.76$ &
    $53.2$ & $28.2$ & $23.8$ \\[0.1 cm]
$ 90$  & $0.0^{+9.7}_{-0.0}$ & $0.55$ &
    $28.9$ & $29.0$ & $18.0$ \\[0.1 cm] 
$100$  & $0.0^{+7.6}_{-0.0}$ & $0.41$ &
    $22.8$ & $27.2$ & $16.8$ \\[0.1 cm] 
$110$  & $0.0^{+6.3}_{-0.0}$ & $0.30$ &
    $18.7$ & $30.1$ & $17.1$ \\[0.1 cm] 
$120$  & $0.0^{+5.9}_{-0.0}$ & $0.20$ &
    $17.6$ & $25.0$ & $16.0$ \\[0.1 cm] 
$130$  & $0.0^{+5.5}_{-0.0}$ & $0.12$ &
    $16.7$ & $38.5$ & $19.7$ \\[0.1 cm] 
$140$  & $0.0^{+5.1}_{-0.0}$ & $0.06$ &
    $15.3$ & $34.5$ & $17.2$ \\[0.1 cm] 
\hline
\hline
\end{tabular}
\end{center}
\end{table} 
\end{small}

  \begin{figure}[t]
    \centerline{
    \hskip 0.2cm
     \epsfig{file=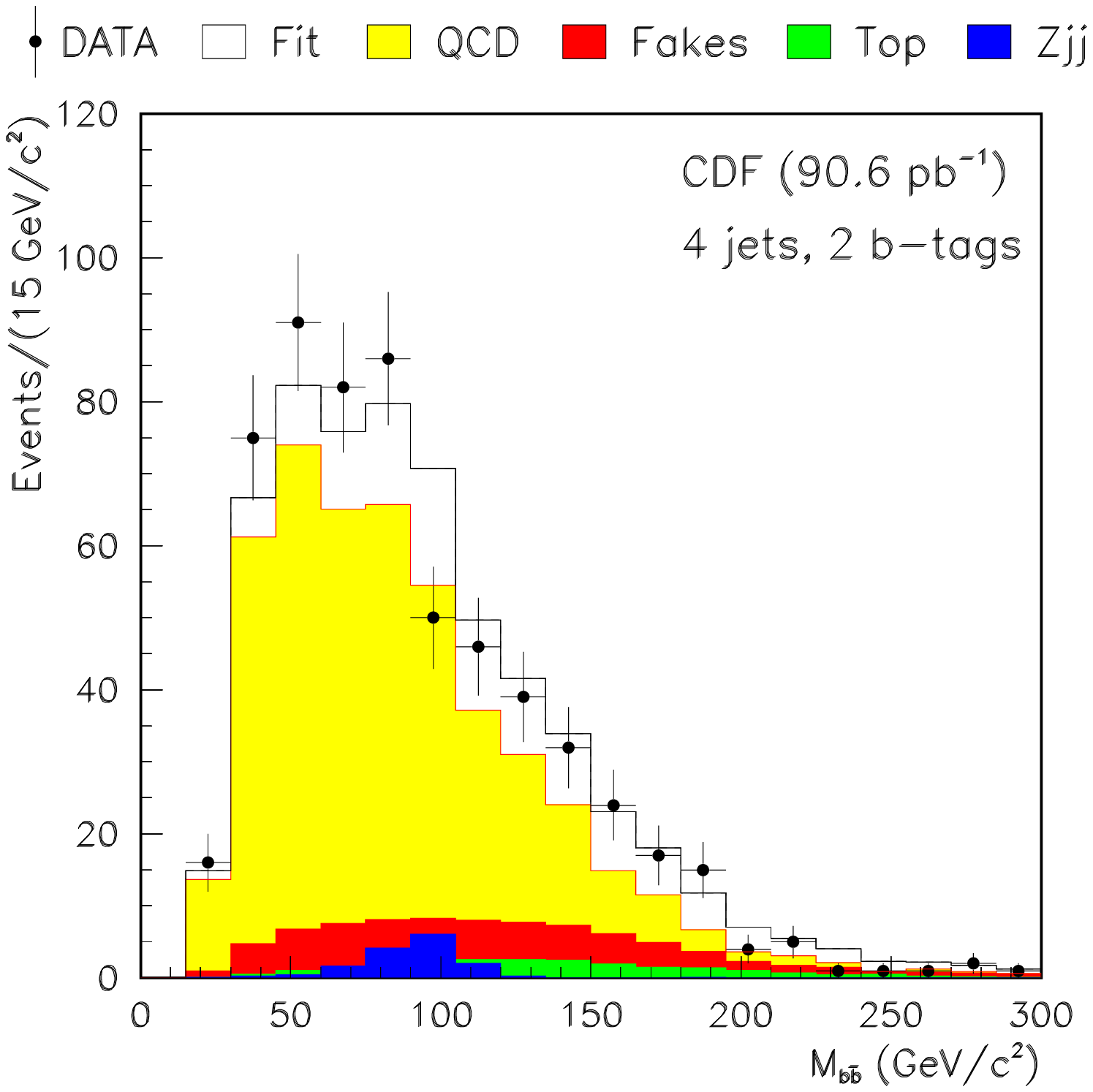,width=8cm}
     \epsfig{file=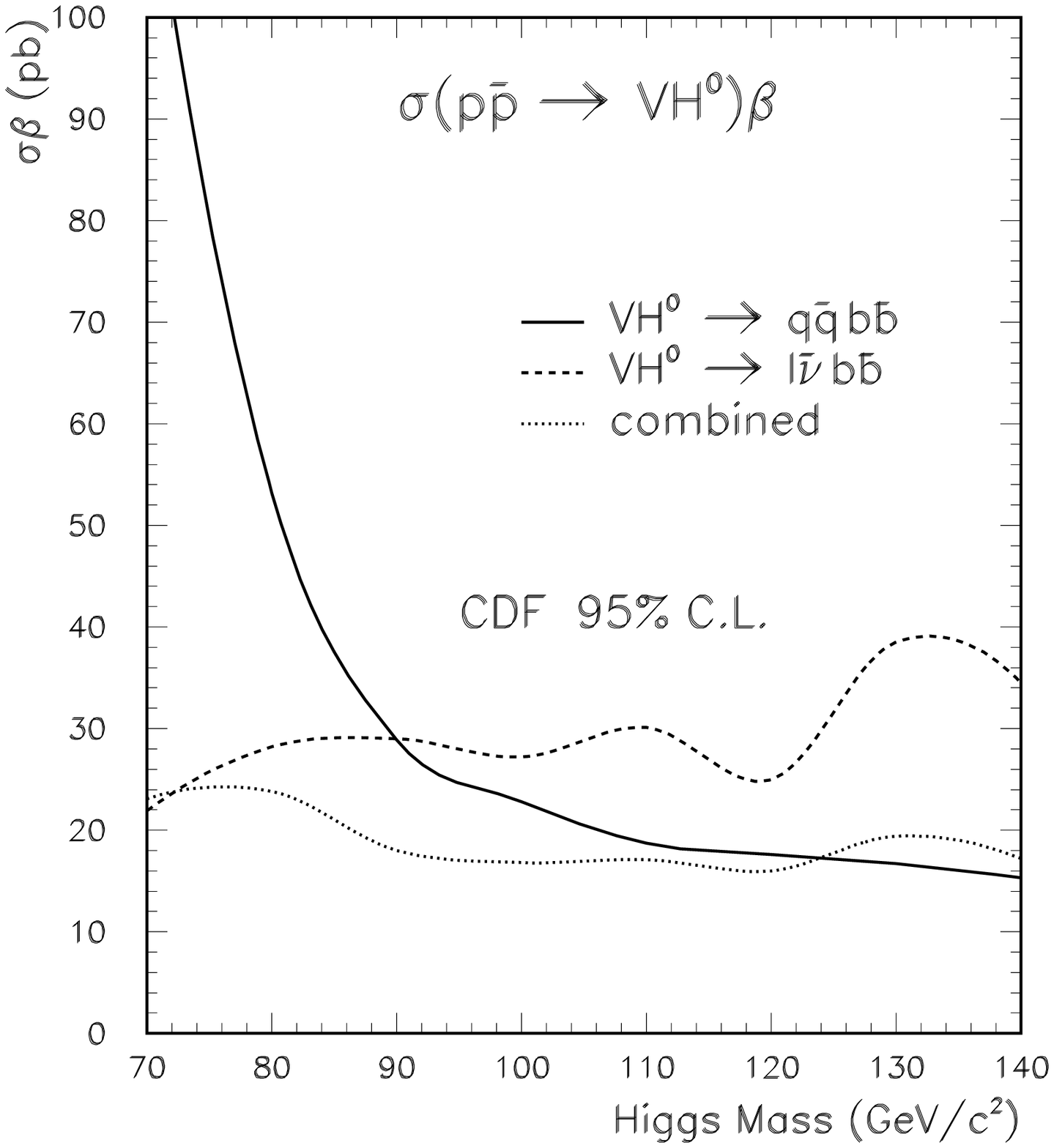,width=8cm}}
    \caption{(Left) 
The measured invariant mass distribution from double-tagged events (points) 
compared to the fit prediction.
The solid line is the sum of the QCD, fakes, $t\bar{t}$, and
$Z$ + jets components.
             (Right) 
The CDF 95\% C.L.  upper 
limits on $\sigma(p\bar{p} \rightarrow VH^0) \cdot \beta$ where 
$\beta = {\cal B}(H^0 \rightarrow b\bar{b})$.}
    \label{fig2}
  \end{figure}

The data sample was recorded with a trigger
which requires four or more clusters of contiguous 
calorimeter towers, each with transverse energy $E_T \ge 15$ GeV, and 
a total transverse energy $\sum E_T \ge 125$ GeV. 
Offline, events are required to have four or more jets with uncorrected
$E_T > 15$ GeV and $|\eta| < 2.1$. 
After this initial selection the sample contains $207,604$ events.
In addition, we require that at least two among the four highest-$E_T$ 
jets in the event tagged by the SECVTX algorithm.

There are 764 events with four or more jets and two or more $b$-tags.
In these events, only the four highest-$E_T$ jets are considered
for the mass reconstruction: the two highest-$E_T$ $b$-tagged 
jets are assigned to the Higgs boson, and the other two to
the vector boson. The $b\bar{b}$ invariant mass distribution in signal 
events contains
a Gaussian core with a sigma of $\sim 0.14\times M_{H^0}$.
The tails of the distribution are dominated by the cases (25-30\%) 
where the jet assignment in the mass reconstruction is incorrect. 
In most of these cases, one of the jets assigned to the Higgs is a heavy 
quark jet from the decay of the $V$ boson.

The challenge of this analysis is to understand the sample composition.
The main source of background events is QCD 
heavy flavor production. The heavy flavor content of QCD hard 
processes has been modelled with PYTHIA.
We generated all QCD jet production channels
and retained the events that contained a heavy quark produced either
in the hard scattering or in the associated radiation process.
Events with a heavy quark are conventionally classified in three groups: 
direct production, gluon splitting, and flavor 
excitation. Direct production events are characterized by a high value of 
the invariant mass, $M_{b\bar{b}}$, and a low value of the 
transverse momentum of the $b\bar{b}$ system, $p_T({b\bar{b}})$. 
The same is true for flavor excitation events. 
The kinematics of final state gluon splitting
events favor a relatively smaller invariant mass value and a large
$p_T({b\bar{b}})$, since both jets tend to be emitted
along the same direction.
In this plane, the
Higgs signal shows a greater tendency to large
$M_{b\bar{b}}$ and $p_T({b\bar{b}})$ values. 
A cut on $p_T({b\bar{b}})\ge 50$ GeV/$c$ is $\sim 80\%$ 
efficient for the signal and strongly discriminates
against direct production and flavor excitation of heavy quarks.
After the $p_T(b\bar{b})$ requirement is applied to the data
589 events remain.

Other backgrounds are $t\bar{t}$ production, $Z$ + jets 
events with $Z\rightarrow b\bar{b}/c\bar{c}$
and fake double-tags. The first two are estimated from Monte Carlo 
and the last one from data. 
Using the CDF measured $t\bar{t}$ production cross section and
a top quark mass of $M_t = 175$ GeV/$c^2$, HERWIG 
predicts $26\pm 7$ $t\bar{t}$ events in the data, 
after trigger, kinematic and $b$-tag requirements. 
The same generator predicts $17 \pm 4$ $Z$ + jets background events.
Fake double-tags are defined as events in which at least one of the two tagged 
jets contains a false secondary vertex in a light quark or gluon jet.
Fake tag probabilities are parameterized by measuring 
in several inclusive jet data samples the proportion of jets 
in which a secondary vertex is reconstructed on the wrong
side of the primary vertex with respect to the jet 
direction~\cite{secvtx}.
The current data set is estimated to contain
$89\pm 11$ fake double-tag events.
Finally, other minor sources of background 
account for less than 1\% of the total number of events, have a broad 
invariant mass distribution, and are neglected in the final fit.

The total signal detection efficiency is defined as the product of the 
trigger efficiency, the kinematical and geometrical acceptances, the double 
$b$-tagging e\-ffi\-cien\-cy, the $p_T(b\bar{b})$ cut efficiency,
and the $V$ hadronic branching fractions. 
The total efficiency increases linearly from 
$0.6\pm 0.1\%$ to $2.2\pm 0.6\%$ for Higgs masses ranging from 70 GeV/$c^2$ 
to 140 GeV/$c^2$.

The shape of the observed $b$-tagged dijet invariant mass distribution is fit,
using a binned maximum-likelihood method, to a combination of  
signal, fake double-tag events, and QCD, $t\bar{t}$ and 
$Z$ + jets backgrounds. 
The QCD and signal normalizations are left free in the fit while
the normalizations of the $t\bar{t}$, $Z$ + jets and fakes are 
constrained by Gaussian functions to their expected values and 
uncertainties. 

The fit yields $\sigma_{VH^0} \cdot \beta = 44 \pm 42$ pb 
for $M_{H^0} = 70$ GeV/$c^2$,
statistically compatible with zero signal.
For larger masses, zero signal contribution is preferred.
Table~\ref{tab1} shows the result of the fits  
as a function of the Higgs mass. Figure~\ref{fig2} (left) shows the 
$b$-tagged dijet invariant mass distribution for the data 
compared to the results of the fit for $M_{H^0} \ge 80$ GeV/$c^2$.

Since the observed distribution is consistent with standard model
background expectations, we place limits on $p\bar{p}
 \rightarrow VH^0$  production.  
Systematic uncertainties on the 95\% C.L. limits
arise from luminosity, jet energy scale, double $b$-tagging efficiencies, 
QCD ra\-dia\-tion, limited Monte Carlo statistics, 
and background normalizations and shapes.
The total systematic uncertainty is in the range $26\%-30\%$.
The 95\% C.L. limits are summarized in
Table~\ref{tab1} and Figure~\ref{fig2} (right). The resulting bounds
fall rapidly from 117 pb at $M_{H^0} = 70$ GeV/$c^2$ to values between 15 and
20 pb for $M_{H^0} > 105$ GeV/$c^2$.

\subsection{Combined results}

To combine the two results presented above,
the data from both channels were then fitted simultaneously~\cite{higgs2}. 
Correlations
between systematic uncertainties due to luminosity, QCD radiation, 
and $b$-tagging efficiency were taken into account. 
All other systematic uncertainties were considered uncorrelated. 
The 95\% C.L. limits range from 16 to 24 pb and are shown 
in Table~\ref{tab1} and Figure~\ref{fig2} (right).  

\section{Technicolor}

Now we turn to the second mechanism of EWSB discussed in the 
Introduction~\cite{lq3}.
Particularly interesting from the present experimental point of view
\cite{sgtures,ramana} are
TC models containing a technifamily, i.e. a set of technifermions 
with the same structure and 
quantum numbers of a complete standard model generation of 
quarks and leptons, and carrying an additional TC quantum number.
By convention, technifermions which are 
color-triplets of ordinary QCD are called techniquarks, and color-singlet
technifermions are called technileptons.
The particle spectrum of these  models includes color-singlet, -triplet and 
-octet technipions.
The technipions ($\pi_{\dinkyT}$) decay via ETC interactions. 
Since these are also responsible for the fermion masses, technipions 
are expected to have Higgs-boson-like couplings to ordinary fermions,
i.e. to decay preferentially to third-generation quarks and leptons. 
In particular, the color-triplet technipions are an example of scalar 
third-generation leptoquarks ($\pi_{\dinkyLQ}$).
In this section, we use the results of a search for third-generation 
leptoquarks by CDF, in order to explore TC models 
containing a technifamily. Other experimental constraints on these
models come from precision electroweak measurements at LEP~\cite{lep-tc,Zbb},
and from measurements of the $b \to s \gamma$ decay rate~\cite{cleo}.

\begin{small}
\begin{table}[t]
\begin{center}
\caption{The 95\% confidence level upper limits on
the leptoquark (color-triplet technipion) production cross section 
times branching ratio squared as a function of 
$M(\pi_{\dinkyLQ})$ and $M(\rho_{\dinkyT})$, for $\Delta M = 50$ GeV$/c^2$. 
Numbers are given in pb.}
\label{tab:dm50}
\begin{tabular}{cccccccccccc}
\hline
\hline
$M(\pi_{\dinkyLQ})$ &
\multicolumn{11}{c}{$M(\rho_{\dinkyT})$ (GeV$/c^2$)}\\
\multicolumn{1}{c}{(GeV$/c^2$)}
    & 200 & 250 & 300 & 350 & 400 & 450 & 500 & 550 & 600 & 650 & 700\\
\hline
100 &12.7 & 9.8 & 8.2 & 7.4 & 7.2 & 7.7 & 8.5 & 9.4 & 9.8 &10.0 &10.2 \\
125 &     & 6.4 & 5.3 & 4.6 & 4.1 & 3.9 & 3.9 & 4.1 & 4.5 & 4.8 & 5.0 \\
150 &     &     & 4.7 & 4.1 & 3.6 & 3.3 & 3.1 & 3.0 & 3.1 & 3.2 & 3.5 \\
175 &     &     &     & 3.7 & 3.3 & 3.1 & 2.9 & 2.7 & 2.6 & 2.6 & 2.7 \\
200 &     &     &     &     & 3.4 & 3.0 & 2.8 & 2.5 & 2.3 & 2.2 & 2.1 \\
225 &     &     &     &     &     & 2.9 & 2.7 & 2.5 & 2.3 & 2.2 & 2.1 \\
250 &     &     &     &     &     &     & 2.8 & 2.5 & 2.3 & 2.2 & 2.1 \\
275 &     &     &     &     &     &     &     & 2.5 & 2.3 & 2.2 & 2.0 \\
300 &     &     &     &     &     &     &     &     & 2.3 & 2.2 & 2.0 \\
\hline
\hline
\end{tabular}
\end{center}
\end{table}
\end{small}

  \begin{figure}[tb]
    \centerline{
     \epsfig{file=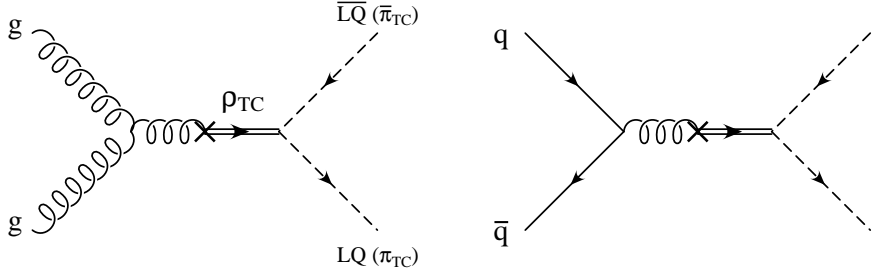,width=11.5cm}}
    \caption{
The resonant production of technipion pairs.
The technirho couples directly to the gluon via vector-meson-dominance 
enhancing s-channel production.}
    \label{fig:resonlq}
  \end{figure}
  
  \begin{figure}[t]
    \centerline{
    \hskip -0.4cm
     \epsfig{file=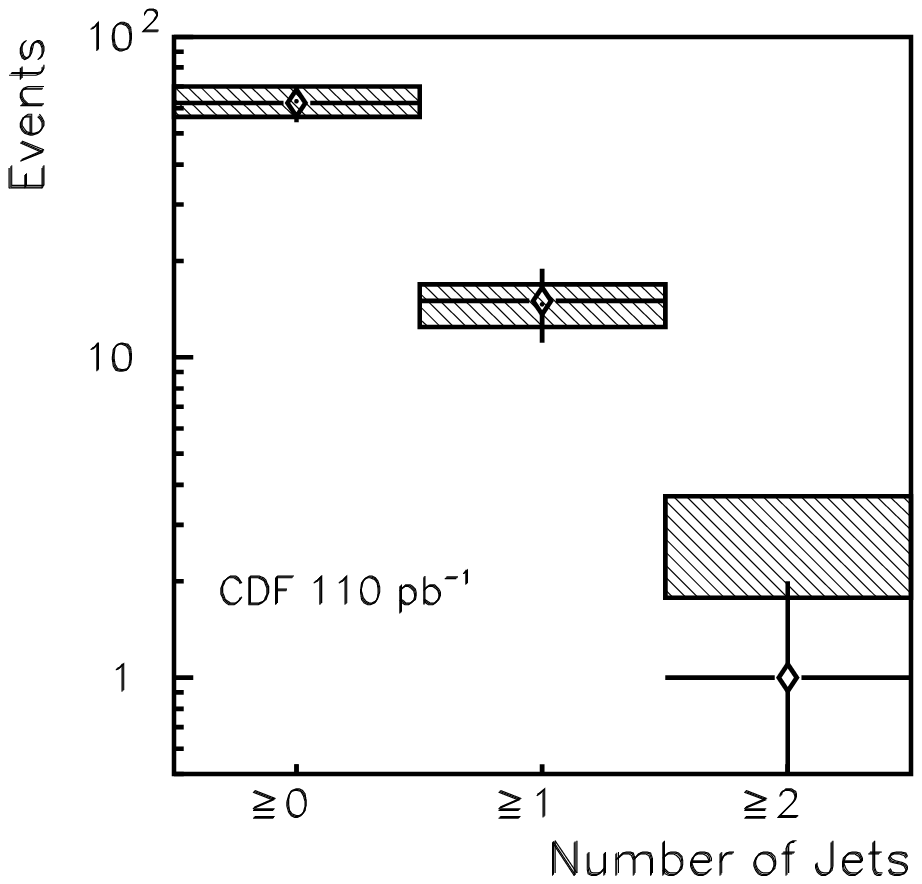,width=8cm}
    \hskip -0.5cm
     \epsfig{file=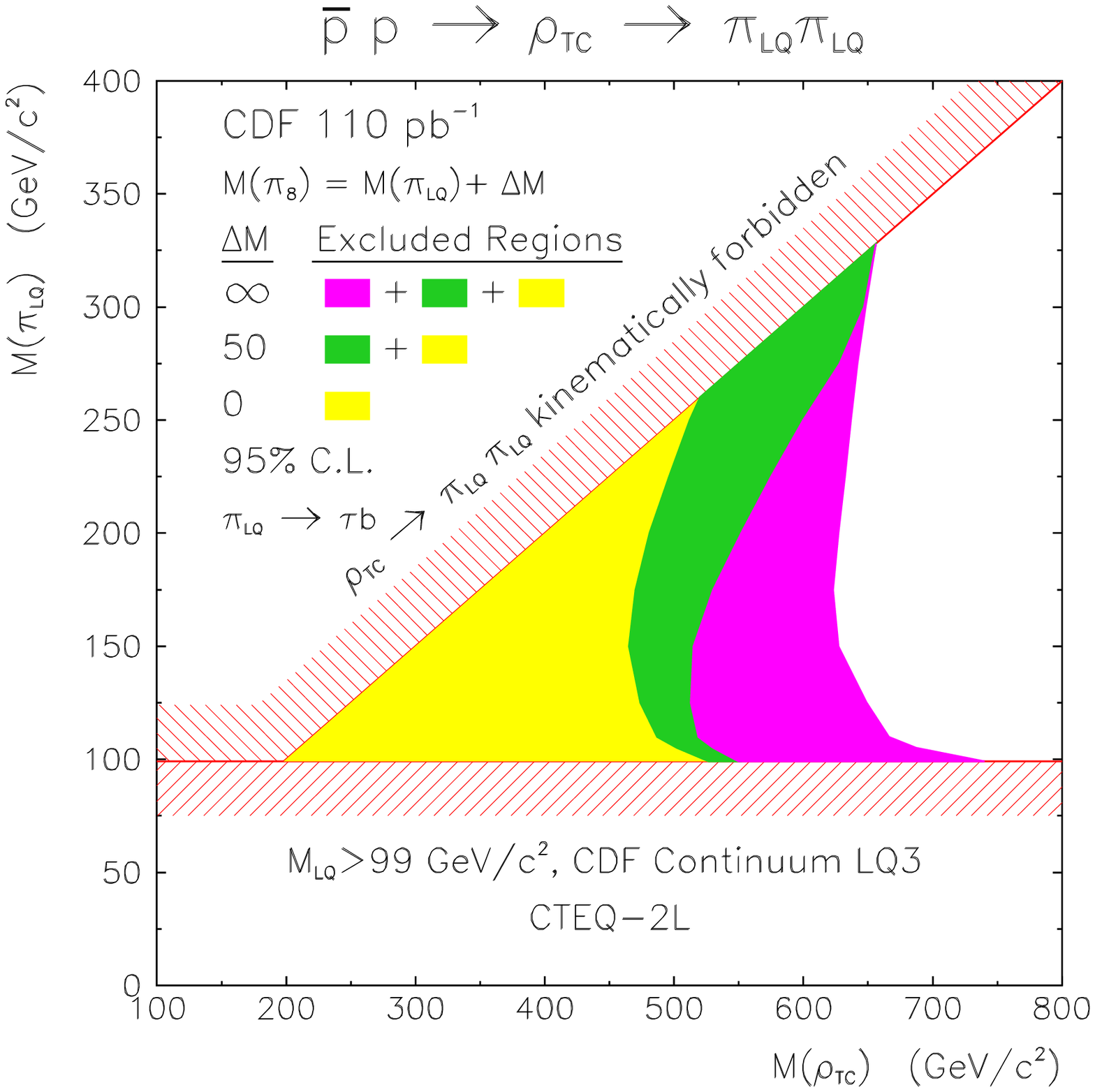,bbllx=0pt,bblly=140pt,bburx=550pt,bbury=680pt,
                        width=8cm,clip=}}
    \caption{(Left) 
The inclusive jet multiplicity in $\tau^+\tau^-$ candidate events (diamonds)
compared to standard model expectations.
             (Right) 
The 95\% C.L. exclusion regions in the 
$M(\pi_{\dinkyLQ}) - M(\rho_{\dinkyT})$ plane.
The three shaded areas correspond (from left to right) to technipion
mass splitting values of 0, 50 GeV$/c^2$ and $\infty$, respectively.}
    \label{fig:exclusion}
  \end{figure}

Let's consider technicolor models containing a family of color-singlet
technileptons and color-triplet techniquarks. 
In these models,
there is a color-octet vector resonance, called technirho ($\rho_{\dinkyT}$),
with the quantum numbers of the gluon.
Leptoquarks are assumed to be pair produced via gluon-gluon fusion 
and $q \bar q$ annihilation.
In $q \bar{q}$ and $gg$ collisions, the $\rho_{\dinkyT}$ couples to the gluon 
propagator enhancing s-channel reactions (Fig.~\ref{fig:resonlq}),
analogously to the vector-meson-dominance description of the process
$e^+ e^- \to \pi^+ \pi^-$~\cite{VMD}.
Two decay modes may exist for the technirho~\cite{ramana}:
$\rho_{\dinkyT} \to q\bar{q}, gg$ and 
$\rho_{\dinkyT} \to \pi_{\dinkyT} \bar{\pi}_{\dinkyT}$.
If the $\rho_{\dinkyT}$ mass is less than twice the 
$\pi_{\dinkyT}$ mass, only the 
$q\bar{q}, gg$ decay mode is possible, resulting in resonant dijet 
production. A search result for the dijet signal of $\rho_{\dinkyT}$ has 
already been reported by CDF. The CDF-measured dijet mass
spectrum excludes $\rho_{\dinkyT}$ with masses in the range 
$260 < M(\rho_{\dinkyT}) < 480$ GeV$/c^2$ at the 95\% C.L. \cite{harris}.
If the $\rho_{\dinkyT}$ mass is larger than twice the 
$\pi_{\dinkyT}$ mass, the $\rho_{\dinkyT}$ 
decays preferentially into $\pi_{\dinkyT}$ pairs. 

The technipion spectrum of the technifamily model was estimated 
in~\cite{peskin,ramana}. It contains color-singlet, 
-triplet and -octet ($\pi_8$) technipions. The 
octets are heavier than the triplets, and these are heavier
than the singlets. We make the simplifying assumption  that 
there is no mass splitting among the different 
octet and triplet technipions. 
We consider the class of color-triplet technipions decaying via 
$\pi_{\dinkyLQ} \to \bar{b} \tau^-$ ($\bar{\pi}_{\dinkyLQ} \to b\tau^+$)
with branching fraction $\beta$.

The leading-order leptoquark pair production cross section depends only on 
the technirho mass ($M(\rho_{\dinkyT})$), the 
leptoquark mass ($M(\pi_{\dinkyLQ})$), 
and the technirho width ($\Gamma(\rho_{\dinkyT})$). $M(\pi_{\dinkyLQ})$ and 
$M(\rho_{\dinkyT})$ are treated as independent free parameters. 
$\Gamma(\rho_{\dinkyT})$ 
can be calculated as a function of four more basic quantities, 
$\Gamma(\rho_{\dinkyT}) = \Gamma(M(\rho_{\dinkyT}),M(\pi_{\dinkyLQ}),
\Delta M,N_{\dinkyTC})$,
where $\Delta M = M(\pi_8) - M(\pi_{\dinkyLQ})$, and $N_{\dinkyTC}$
is the number of technicolors. We consider $M(\rho_{\dinkyT})$, 
$M(\pi_{\dinkyLQ})$, 
$\Delta M$, and $N_{\dinkyTC}$ as the four continuous parameters of the
theory. We set limits in the $M(\pi_{\dinkyLQ}) - M(\rho_{\dinkyT})$ plane.
We probe the dependence of the production cross section on 
$\Gamma(\rho_{\dinkyT})$ by fixing
$N_{\dinkyTC} = 4$, while allowing $\Delta M$ to take one expected
and two limiting values. ETC and QCD corrections to $M(\pi_8)$ 
and $M(\pi_{\dinkyLQ})$ are responsible for $\Delta M$, analogously to
the QED corrections to $M(\pi^o)$ and $M(\pi^\pm)$.  $\Delta M$ is expected 
to be around 50 GeV$/c^2$ \cite{ramana}. 
We take $\Delta M = 0$ and $\Delta M = \infty$ as two extreme values. 
The resulting variation in $\Gamma(\rho_{\dinkyT})$ could also have 
been obtained changing $N_{\dinkyTC}$ by a factor of 4, for a fixed 
$\Delta M = 50$ GeV$/c^2$.
 
The experimental signature considered is $\tau^+ \tau^-$ plus two jets 
in the final state, in the case where one $\tau$ decays leptonically and 
the other decays hadronically.  
The analysis selects a 110~pb$^{-1}$ 
data set containing an isolated electron or muon in the region 
$|\eta| < 1$ with $\pt > 20$ GeV$/c$~\cite{CDFcoo}, and an isolated, 
highly-collimated hadronic jet consistent with a hadronic tau decay.
Hadronic $\tau$ candidates ($\tau$-jets) are selected from
jets that have an uncorrected total transverse energy of 
\mbox{$\et > 15 \gev$} in the
region \mbox{$| \eta | < 1$}.
The associated charged particles with \mbox{$p_T >$ 1 GeV/c}
in a 10$^{\rm o}$ cone around the jet direction
must satisfy the following requirements:
($i$)
the $\tau$-jet must have one or three charged particles;
($ii$)
if there are three, the scalar sum $\pt$
must exceed 20 GeV/c
and the invariant mass must be smaller than 2 GeV/c$^2$;
and ($iii$)
the leading charged particle must have $\pt > 10 \gev$/c and
must point to an instrumented region of the calorimeter.
The efficiency of the $\tau$-jet identification criteria
grows from 32\% for $\tau$-jets in the range $15 < \et < 20 \gev$ to a 
plateau value of 59\% for $\et > 40 \gev$.
Isolated $\tau$-jets must have
no charged particles with $\pt >$ 1 \gev/c in the annulus between
10$^{\rm o}$ and 30$^{\rm o}$ around the jet axis.
Events where the high-$\pt$ lepton is consistent with originating
from a $Z \to ee$ or $Z \to \mu \mu$ decay are removed.
In addition, the analysis uses the
missing transverse energy characteristic of neutrinos from tau 
decays. 
The requirement \mbox{$\Delta \Phi < 50^{\rm o}$}, where
\mbox{$\Delta \Phi$} is the azimuthal separation
between the directions of the missing transverse energy $\met$ and the
lepton,
distinguishes $\tau^+ \tau^-$ events from backgrounds such as
$W$ + jets.
Figure~\ref{fig:exclusion} (left) shows the jet multiplicity
in  $\tau^+ \tau^-$ candidate events. The agreement with the
standard model background prediction is excellent.
Finally, two or more jets with $E_T >$ 10 GeV and $|\eta| <$ 4.2, 
assumed to originate from $b$ quark hadronization, are required.
One leptoquark pair 
candidate event survives these selection criteria. The observed yield is 
consistent with the $2.4^{+1.2}_{-0.6}$ expected background events from 
standard model 
processes, dominated by $Z \to \tau \tau + $ jets production ($2.1 \pm 0.6$) 
with the remainder from diboson and $t \bar{t}$ production~\cite{lq3}.

The detection efficiencies for the signal are determined using a full 
leading-order 
matrix element calculation for technipion pair production 
(continuum, resonant, and interference terms are included)~\cite{ramana} 
and embedded in the PYTHIA Monte Carlo program
to model the full $p \bar{p}$ event structure. 
The generated events are passed through a detector simulation program 
and subjected to the same search requirements as the data.
The total efficiency increases from 0.3\% for $M(\rho_{\dinkyT}) =$ 200
GeV$/c^2$ and $M(\pi_{\dinkyLQ}) =$ 100 GeV$/c^2$, to 1.8\%
for $M(\rho_{\dinkyT}) =$ 700 GeV$/c^2$ and 
$M(\pi_{\dinkyLQ}) =$ 300 GeV$/c^2$.
The systematic errors in the efficiencies were estimated as described in 
\cite{lq3}, including uncertainties in the modelling of
gluon radiation, in the calorimeter energy scale, in the dependence
on renormalization scales, and in the luminosity measurement.
They range from 15\% for $M(\rho_{\dinkyT}) =$ 200 GeV$/c^2$ and 
$M(\pi_{\dinkyLQ}) =$ 100 GeV$/c^2$, to 10\%
for $M(\pi_{\dinkyLQ}) \ge$ 125 GeV$/c^2$.

We place limits on the leptoquark pair production 
cross section times branching ratio squared within the framework of the 
technicolor model described above.
Table~\ref{tab:dm50} lists the leptoquark 95\% confidence level upper 
limits on the production cross section times branching ratio squared as a 
function of 
$M(\pi_{\dinkyLQ})$ and $M(\rho_{\dinkyT})$, for $\Delta M = 50$ GeV$/c^2$. 
These numbers differ by at most 1 pb from the 
corresponding limits for $\Delta M = 0$ and $\Delta M = \infty$ 
when $M(\pi_{\dinkyLQ}) < 175$ GeV$/c^2$.  For larger values
of $M(\pi_{\dinkyLQ})$ the differences are negligible.  
Comparing to the theoretical expectations for $\sigma(p \bar{p} \to 
\pi_{\dinkyLQ} \bar{\pi}_{\dinkyLQ}) \cdot \beta^2$,
we place bounds in the $M(\pi_{\dinkyLQ}) - M(\rho_{\dinkyT})$ plane.
Figure~\ref{fig:exclusion} (right) shows the
95\% C.L. mass exclusion regions.
The upper part of the plot corresponds to the kinematically forbidden region
where $M(\rho_{\dinkyT}) < 2 M(\pi_{\dinkyLQ})$. The bottom region
is the exclusion area from the continuum leptoquark analysis,
$M(\pi_{\dinkyLQ}) \geq 99$ GeV$/c^2$~\cite{lq3}.
The three shaded areas from left to right correspond to technipion
mass splitting values of $\Delta M = 0$, 50 GeV$/c^2$ and $\infty$,
respectively.  
Although more information is presented in Figure~\ref{fig:exclusion},
it is useful to summarize our technirho excluded region using a single
number. For $\Delta M = 0$ (50, $\infty$) and 
$M(\pi_{\dinkyLQ})  < M(\rho_{\dinkyT})/2$,
we exclude color octet technirhos with mass less than 465 (513, 622) 
GeV$/c^2$ at 95\% confidence level.

\section{Discussion and Prospects}

We have described several attempts to explore the phenomenology of the EWSB 
sector at the Tevatron. A first kind of analyses tried to reconstruct 
$b\bar b$ resonances. From the experimental point of view, this involved 
the study and development of a number of techniques of broad interest 
for the hadron collider experiments of the next decade. These include 
jet spectroscopy, $b$-tagging algorithms, and quantitative evaluation of 
challenging new backgrounds as QCD production of heavy quarks in a
multijet environment. 

No signal has been found yet, and the results have been used to constrain 
the production of a light standard model Higgs.
The Higgs searches have not provided yet any lower Higgs mass bound. The 
sensitivity of the present search is limited by statistics to a cross 
section approximately two orders of magnitude larger than the predicted 
cross section for standard model Higgs production~\cite{theo1}. It also 
should be noted that, because these 
limits were derived from a shape fit, they only apply to a very 
restricted region of parameter space in the minimal supersymmetric 
extension of the standard model.
 
For the next Tevatron run CDF hopes for an approximately twenty-fold 
increase in the total integrated luminosity and a factor of two improvement 
in the double $b$-tagging efficiency. D0 will be as good as CDF in this 
respect. However, this is still insufficient
to reach say a 120 GeV$/c^2$ Higgs mass, unless the total detection 
efficiency be improved by one order of magnitude. The viability of this 
improvement is at present under study~\cite{john}. Plans include the 
installation of dedicated Higgs triggers, systematic inclusion of 
all relevant channels 
(notably $ZH^0$ production, with $Z \to \ell \ell$ and $Z \to \nu \nu$), 
finer mass resolution, and additional, more efficient 
selection algorithms (e.g., neural networks).

Another analysis described a virtually background-free search of resonances 
decaying into $\tau$ leptons. The four objects appearing in the selected events
can be arranged in two ways. Looking for $\tau$-jet leptoquark resonances
the analysis can be used to directly constrain TC models including one
technifamily. The expected factor $\times$20 in luminosity will help
to push the technirho mass limits closer to the TeV region. 
Interpreted as containing a $\tau^+ \tau^-$ resonance, the same events
can be used to test effectively the large tan($\beta$) region of the
SUSY extension of the standard model~\cite{drees}.

In conclusion, we are just at the very beginning of this kind of physics
at hadron colliders. Most importantly we need data to look at, and for
that reason the CDF and D0 Collaborations are looking forward to Run 2
with great anticipation.

\section*{Acknowledgments}
I would like to thank Prof. B. Adeva and collaborators for the stimulating 
atmosphere of the meeting and warm hospitality.


\begin{thebibliography}{99}
\bibitem{CDFcoo} 
In the CDF coordinate system, $\phi$ and $\theta$ are the
azimuthal and polar angles with respect to
the proton beam direction.
The pseudorapidity $\eta$
is defined as $-\ln[\tan (\theta/2)]$.
The transverse momentum of a particle
is $\pt = p \sin \theta$.
The analogous quantity using calorimeter energies is
called the transverse energy $\et$.
The difference between the vector sum of all the
transverse energies and zero is the missing transverse energy $\met$.
\end{thebibliography}
\end{document}